\documentclass[aps,prl,twocolumn,showpacs]{revtex4}
\usepackage{epsfig}
\usepackage{times}
\begin{document}

\title{Percolation threshold determines the optimal population density for public cooperation}

\author{Zhen Wang,$^{1,2}$ Attila Szolnoki,$^3$ Matja{\v z} Perc$^4$}
\affiliation{$^1$School of Physics, Nankai University, Tianjin 300071, China\\
$^2$Department of Physics, Hong Kong Baptist University, Kowloon Tong, Hong Kong\\
$^3$Institute for Technical Physics and Materials Science, Research Centre for Natural Sciences, Hungarian Academy of Sciences, P.O. Box 49, H-1525 Budapest, Hungary \\
$^4$Faculty of Natural Sciences and Mathematics, University of Maribor, Koro{\v s}ka cesta 160, SI-2000 Maribor, Slovenia}

\begin{abstract}
While worldwide census data provide statistical evidence that firmly link the population density with several indicators of social welfare, the precise mechanisms underlying these observations are largely unknown. Here we study the impact of population density on the evolution of public cooperation in structured populations, and find that the optimal density is uniquely related to the percolation threshold of the host graph irrespective of its topological details. We explain our observations by showing that spatial reciprocity peaks in the vicinity of the percolation threshold, when the emergence of a giant cooperative cluster is hindered neither by vacancy nor by invading defectors, thus discovering an intuitive yet universal law that links the population density with social prosperity.
\end{abstract}

\pacs{87.23.Ge, 89.75.Fb}
\maketitle

When performing his seminal experiments on the behavior of rats under crowded conditions, ethologist John B. Calhoun found that too high population densities may induce a variety of destructive conditions, ranging from infant cannibalism over excessive aggression to increased mortality at all ages \cite{calhoun_sa62}. These observations became known as the ``behavioral sink'', and it was later confirmed that similar, although not quite as savage and somewhat more subtle, effects of overcrowding can be observed not just by rodents, but also by primates \cite{judge_ab97} and humans \cite{galle_s72}. Although there is some disagreement amongst sociologists as to how much population density actually affects human behavior \cite{freedman_jesp75, baldassare_abs75} and what are its implications for welfare participation \cite{rank_d93}, fact is that World maps, depicting increasing population density over a certain point on one side and decreasing social welfare indexes on the other, as well as freely available census data, dispel all doubts concerning their relatedness.

Cooperation in sizable groups is a particularly interesting social phenomenon \cite{nowak_11}, as it is arguably crucial for the remarkable evolutionary success of the human species. While the origins of human cooperation are most frequently attributed to between-group conflicts \cite{bowles_11} and alloparental care \cite{hrdy_11}, mechanisms such as kin and group selection, as well as direct, indirect and spatial reciprocity, are known to facilitate its evolution \cite{nowak_s06}. The public goods game captures succinctly the essential social dilemma related to the evolution of cooperation in groups \cite{sigmund_10}. Players must decide simultaneously whether they wish to contribute to the common pool or not. All individual contributions, for simplicity assumed being equal to one, are then multiplied by a factor $r>1$ to take into account synergetic effects of cooperation, and the resulting amount is divided equally among all group members irrespective of their strategy. Clearly, individuals are tempted to defect, while the group as a whole is best off if everybody cooperates. Failure to harvest the benefits of a collective investment and mindless exploitation of public goods are in fact the key causes for the ``tragedy of the commons'' \cite{hardin_g_s68}.

During the past decade, physics-inspired studies have led to significant advancements in our understanding of the evolution of cooperation, especially related to games on graphs \cite{szabo_pr07} and coevolutionary games \cite{perc_bs10}. Inspired by the seminal paper on spatial reciprocity \cite{nowak_n92b}, for example, scale-free networks have proven optimal for the evolution of cooperation \cite{santos_prl05}, while the dynamical organization of cooperation on complex networks has provided vital insights as to why this is the case \cite{gomez-gardenes_prl07}. Most recently, evolutionary games have also been studied in growing populations \cite{poncela_njp09, melbinger_prl10}, as well as on emergent hierarchical structures \cite{lee_s_prl11}. Of direct relevance for the present study are the early works on disordered environments in spatial games \cite{vainstein_pre01, vainstein_jtb07}, which gave rise to studies clarifying the role of mobility in different evolutionary settings \cite{sicardi_jtb09, helbing_pnas09, cheng_hy_njp10}. It is within the latter works that the impact of population density has been investigated before, primarily in relation to optimization possibilities the empty sites give to success-driven individuals, as determined by means of pairwise interactions with other players.

Playing in a group with other players yields many-body interactions, and their consequences cannot always be understood based on pairwise interactions. Motivated by this possibility, we here depart from games governed by pairwise interactions and focus on the spatial public goods game \cite{szolnoki_pre09c}. We investigate the impact of population density on the evolution of public cooperation by using a square lattice of size $L^2$, where only a fraction $\rho$ of all the nodes is occupied by players while the other nodes are left empty. The random dilution of the lattice is performed only once at the start of the game, and initially every player $x$ is designated either as cooperator ($s_x=C=1$) or defector ($s_x=D=0)$ with equal probability. Monte Carlo simulations are carried out comprising the following elementary steps.

A randomly selected player $x$ acquires its payoff $P_{x}^g$ by playing the public goods games with its existing interaction partners as a member of a $g \in G=1 \dots 5$ group whereby its overall payoff is thus $P_{x} = \sum_g P_{x}^g$. Next, player $x$ chooses one of its nearest neighbors at random, and the chosen co-player $y$ also acquires its payoff $P_{y}$ in the same way. Finally, player $x$ enforces its strategy $s_x$ onto player $y$ with a probability $w(s_x \to s_y)=1/\{1+\exp[(P_{y}-P_{x})/K]\}$, where $K=0.5$ quantifies the uncertainty by strategy adoptions. If player $x$ has no nearest neighbors the whole procedure starts anew without attempting a strategy change. Each Monte Carlo step (MCS) gives a chance for every player to enforce its strategy onto one of the neighbors (if they exist, which at sufficiently small $\rho$ will not be the case) once on average. The average density of cooperators $f_{C}=\rho^{-1} L^{-2}\sum_x s_x$ was determined in the stationary state after sufficiently long relaxation times. Depending on the actual conditions the linear system size was varied from $L=200$ to $1200$ and the relaxation time was varied from $10^4$ to $10^6$ MCS to ensure proper accuracy.

\begin{figure}
\begin{center} \includegraphics[width = 7.5cm]{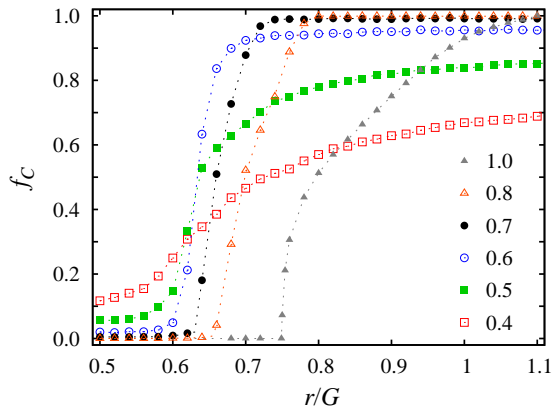}
\caption{\label{motivate}The peculiar dependence of the fraction of cooperators $f_C$ on the normalized synergy factor $r/G$ for different population densities $\rho$ (see legend), as obtained for the square lattice. While decreasing the population density below $1$ facilitates the evolution of public cooperation, there exists a lower bound to $\rho$ below which, for sufficiently high values of $r$, the effect is reversed. This indicates that the population density crucially affects the evolution of public cooperation, but it does so in a non-trivial way.}
\end{center}
\end{figure}

To begin with, it is motivating to examine the evolution of cooperation in dependence on the synergy factor $r$ for different population densities $\rho$. In Fig.~\ref{motivate}, the $\rho=1$ curve recovers the well-known result \cite{szolnoki_pre09c} of cooperator extinction and dominance below $R_{C1}=r/G=0.75$ and above $R_{C2}=r/G=1.1$, respectively. As $\rho$ decreases below one, initially both $R_{C1}$ (the extinction threshold) and $R_{C2}$ (the dominance threshold) decrease, thus indicating that smaller populations densities favorably affect the evolution of public cooperation. Below $\rho=0.6$, however, the positive effect begins to deteriorate, at least partially so. While $R_{C1}$ keeps decreasing, $R_{C2}$ becomes altogether unattainable, \textit{i.e.} cooperators become unable to dominate even at very large values of $r$. These results indicate that $\rho$ plays a key role in games governed by group interactions, invigorating on one hand the previous results obtained for pairwise interactions \cite{vainstein_pre01} as well as the common perception of the importance of population density for social welfare, while on the other inviting a more detailed study as to why this is the case.

\begin{figure}
\begin{center} \includegraphics[width = 7.5cm]{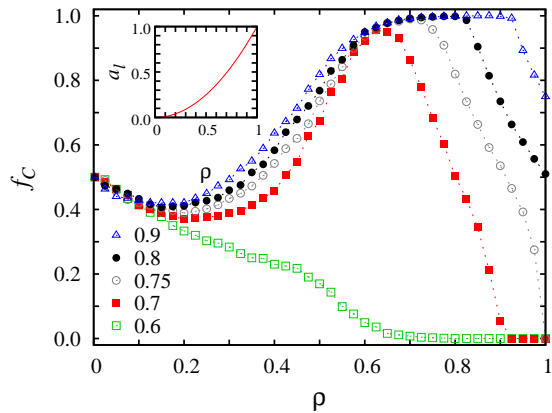}
\caption{\label{sqr} Fraction of cooperators $f_C$ in dependence on the population density $\rho$ for different values or $r/G$ (see legend), as obtained for the square lattice. Inset depicts the corresponding growth of the fraction of active links $a_l$ (occupied nearest neighbor sites) as $\rho$ increases. Cooperators go extinct at $r=R_{C1}=0.75$ if $\rho=1$. The optimal population density where cooperators can dominate even at smaller $r$ is slightly above the percolation threshold, which is  $\pi=0.59$.}
\end{center}
\end{figure}

Results presented in Fig.~\ref{sqr} provide a clearer view of the impact of $\rho$ on $f_C$. In the $\rho \to 0$ limit the majority of players has no neighbors at all (see inset), and hence $f_C$ simply mirrors back the initial state. As $\rho$ increases, the few existing links between players enable defectors to exploit cooperators without having to fear the consequences of spatial reciprocity. Note that for $\rho<0.2$ many players, as well as large portions of the lattice as a whole, will still be disconnected, hence prohibiting cooperators to form compact clusters and utilizing this (spatial reciprocity) to protect themselves against invading defectors. Because of the random initial state, the first strike of defectors will always be successful, regardless of the value of $r$. But further invasions are subsequently hindered by the lack of connections between players utilizing different strategies, and hence at low values of $\rho$ the decay of $f_C$ is universal for all values of $r$. For $\rho>0.2$, however, the outcome of the game becomes dependent on the synergy factor. For low values of $r$ ($r/G=0.6$) the $f_C$ trend simply continues downward as $\rho$ increases, which indicates that new cooperative players simply serve as easy targets for defectors. At higher values of $r$ cooperators are able to utilize the enhanced interconnectedness between them to form compact clusters, while at the same time benefiting from the dilution that prohibits defectors to exploit them with the same efficiency as on a fully populated lattice. Accordingly, $f_C$ peaks at an intermediate (optimal) value of $\rho=\rho_o \approx 0.62$, which is a bit higher but close to the percolation threshold of the square lattice equalling $\pi=0.59$ \cite{stauffer_94}. Upon further increasing $\rho$, the average number of empty sites decays, and accordingly the effective size of groups rises. Since larger groups in general require larger synergy factors to maintain cooperation, $f_C$ remains high at higher $\rho$ only if the value of $r$ is sufficiently large, yet starts falling if $r$ is too small. Thus, the larger the value of $r$ the higher the value of $\rho$ where $f_C$ starts decaying. From this we argue that a population density close to the percolation threshold offers a delicate optimum for the successful evolution of cooperation, where the players are connected enough to transfer the more advantageous mutually beneficial strategy, while simultaneously the lattice is diluted enough to annul free-riders.

\begin{figure}
\begin{center} \includegraphics[width = 7.5cm]{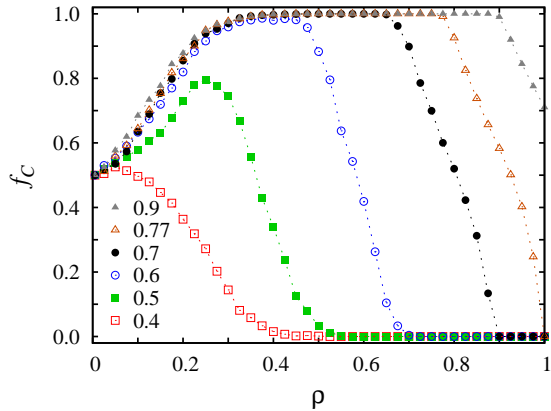}
\caption{\label{sqrext}The optimal population density on the square lattice decreases if strategy imitations are allowed not just between nearest neighbors but also between all the players that are involved in an instance of the public goods game (compare with figure~\ref{sqr}). This is because the extension of the imitation range effectively reduces the percolation threshold. Presented is the fraction of cooperators $f_C$ in dependence on the population density $\rho$ for different values or $r/G$ (see legend). Note that such an extension changes the behavior irrelevantly at $\rho=1$, where $R_{C1}=0.77$.}
\end{center}
\end{figure}

With the aim of closing in on the relevance of the percolation threshold of the interaction graph for the optimal evolution of public cooperation, we alter the public goods game by allowing strategy transfers not just between nearest neighbors, but also between the players that are involved in all the $G=5$ groups. This effectively decreases the percolation threshold as it increases the range of each individual player, while at the same time negligibly affecting the outcome of the game at $\rho=1$.
For $\rho<1$, however, and in particular in the $\rho \to 0$ limit, the interaction graph will be significantly different from the one that is utilized in the standard version of the public goods game. Indeed, due to the significantly lower percolation threshold, the initial decay of $f_C$ as the population density exceeds zero is altogether missing, as can be observed by comparing results presented in Figs.~\ref{sqr} and \ref{sqrext}. More to the point, results in Fig.~\ref{sqrext} support the conclusion that the population density close to the percolation threshold is decisive for a successful evolution of public cooperation. Note that for $r/G=0.5$ the cooperation density peaks at an intermediate value of $\rho$, which in agreement with a lower percolation threshold of the considered lattice occurs at a likewise lower $\rho=\rho_o$. For the same reason the downfall of $f_C$ for higher values of $r/G$ as $\rho \to 1$ is somewhat delayed if compared to the traditional version of the game.

\begin{figure}
\begin{center} \includegraphics[width = 7.5cm]{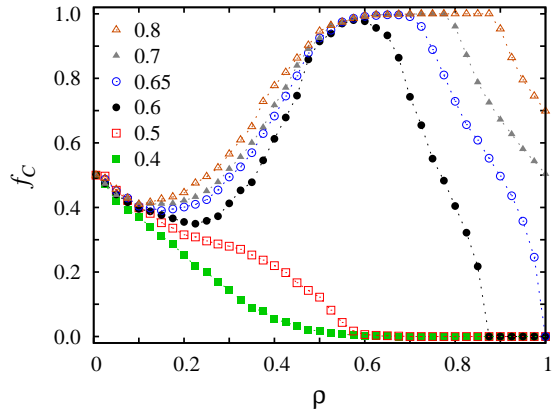}
\caption{\label{tri}
Fraction of cooperators $f_C$ in dependence on the population density $\rho$ for different values or $r/G$ (see legend), as obtained for the triangular lattice. Cooperators go extinct at $r=R_{C1}=0.65$ if $\rho=1$. Like on the square lattice (see Fig.~\ref{sqr}), on the triangular lattice too $f_C$ is independent of $r$ at small values of $\rho$, although the deviations occur sooner (at lower $\rho$) because of the lower percolation threshold. The optimal population density where cooperators can dominate even at smaller $r$ is $\rho_o \approx 0.55$, which is slightly above the percolation threshold ($\pi=0.5$).}
\end{center}
\end{figure}

As another evidence supporting the main message of this work, we show in Fig.~\ref{tri} the same analysis as above for the triangular lattice, whereby as in Fig.~\ref{sqr} strategy imitation is allowed only between nearest neighbors. The triangular lattice has the percolation threshold equal to $\pi=0.5$ \cite{stauffer_94}, while the critical $r/G$ for cooperation extinction on a fully populated lattice is $R_{C1}=0.65$. As Fig.~\ref{tri} illustrates, cooperators can survive or even dominate at smaller $r$ values. Notably, the smallest $\rho$ value where this can happen is slightly above the percolation threshold. At the same time, the independence of $f_C$ on $r$ at small values of $\rho$, as well as the delayed onset of decay of cooperator density for high values of $r$ when $\rho>\pi$, validate the general features that can be understood clearly in terms of the interplay between the evolutionary dynamics and the properties of the interactions graph in both the $\rho \to 0$ and $\rho \to 1$ limit.

\begin{figure}
\begin{center} \includegraphics[width = 8.5cm]{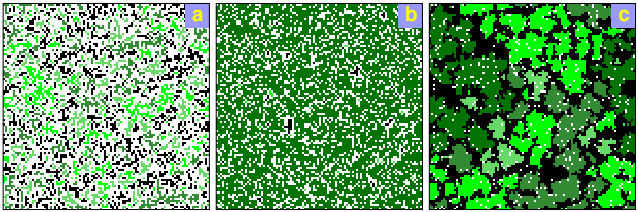}
\caption{\label{perc}Specially prepared snapshots of strategy distributions evidence that only in the proximity of the percolation threshold (panel b) cooperators (green) are able to fully percolate. At lower population densities (panel a) this is prohibited by empty sites (white), while at higher population densities (panel c) percolation is prohibited by defectors (black). Different shades of green (there are only four for clarity) are used for cooperators who belong to different cooperative clusters, \textit{i.e.} cooperators who cannot reach each other by means of nearest-neighbor interactions. Note that the latter is the natural reach of imitation on the square lattice used. Population densities are (a) $\rho=0.4$, (b) $\rho=0.71$ and (c) $\rho=0.95$, while $r/G=0.8$.}
\end{center}
\end{figure}

To understand these results, however, it is necessary to link the evolutionary process itself with percolation. Indeed, there exist compelling evidence that link the extinction of cooperators in the public goods game to the directed percolation universality class \cite{szabo_prl02, helbing_njp10}. But to understand why exactly it is that cooperators are able to percolate optimally even at modest values of $r$ in the vicinity of the percolation threshold, it is instructive to examine characteristic snapshots of strategy distributions at different values of $\rho$, as presented in Fig.~\ref{perc}. While cooperation and defection are always depicted green and black respectively, the shade of green varies depending on which cluster the different cooperators belong to. At low values of $\rho$ there are different shades of green inferable, indicating that while there are clusters of cooperators in existence, these cannot communicate with each other effectively. Remarkably, at high values of $\rho$ the situation is very similar, but for an entirely different reason. While at low values of $\rho$ empty sites (white) disallow cooperators to grow large compact clusters and to communicate with each other, at high values of $\rho$ the defectors are the ones who break up large clusters into isolation and thus diminish the effectiveness of spatial reciprocity between their members. Both ways are equally effective in maintaining a lower level of cooperation, which in panels (a) and (c) is the same. In the vicinity of the percolation threshold, however, there are just enough communication pathways between cooperators to enable their complete percolation (a single giant cooperative domain), yet not as many to sustain the presence of free-riders who could effectively exploit larger groups. Accordingly, spatial reciprocity can be taken full advantage off and cooperation thrives. More precisely, the global density of players should be slightly higher than $\pi$ so that the cooperators who represent only a subset of the whole population can percolate. With this insight, we are thus able to foretell the optimal population density for a given matrix simply by determining its percolation threshold.

Summarizing, we have shown that the percolation threshold of an interaction graph constitutes the optimal population density for the evolution of public cooperation. We have demonstrated this by presenting outcomes of the public goods game on the square lattice with and without an extended imitation range, as well as on the triangular lattice. We have attributed our results to the optimization of spatial reciprocity \cite{nowak_n92b}, the act by means of which connected cooperators share both the production and the benefits of acquired goods. If the population density is below the percolation threshold, vacant sites impede this process by cutting short the communication paths between cooperators. Significantly above it, however, the higher group ``crowdedness'' enables an effective invasion of defectors, which again disrupts reciprocity amongst cooperators by splitting them up into isolated clusters. Presented results offer a new understanding of the impact of population density on social prosperity through the concept of percolation, thus fusing together physics and social science in a mutually rewarding way.

\begin{acknowledgments}
This research was supported by the Hungarian National Research Fund (grant K-73449) and the Slovenian Research Agency (grant J1-4055).
\end{acknowledgments}

\end{document}